\documentclass[12pt,titlepage]{article}

\usepackage{times}
\usepackage{bm}
\usepackage{relsize}

\usepackage[symbol, hang, flushmargin]{footmisc}

\usepackage[utf8]{inputenc}
\usepackage{multicol}
\usepackage{amssymb}
\usepackage{hyperref}
\usepackage[dvipsnames]{xcolor} 
\hypersetup{
    colorlinks=true,
    citecolor=black,     
    urlcolor=RawSienna,
    linkcolor=BlueViolet,
}

\usepackage{sectsty}
\usepackage{enumitem}
\usepackage{tcolorbox}
\allsectionsfont{\normalsize\bfseries}

\usepackage{tabularx}
\usepackage{graphicx}
\usepackage{graphbox}
\usepackage{wrapfig}
\usepackage{caption}

\newtcolorbox{mybox}[2][]{colbacktitle=black!10!white, coltitle=black!80!white, fonttitle=\bfseries,#1}

\usepackage[
    backend=biber,
    style=ieee, 
    citestyle=numeric,
    sorting=none, 
]{biblatex}

\makeatletter
\renewcommand{\maketitle}{
    \begin{center}
    \textbf{Integrating Machine Learning for Planetary Science: Perspectives for the Next Decade}\\
    \vspace{0.2cm}
    \emph{A White Paper to the NRC Planetary Science and Astrobiology Decadal Survey 2023-2032}
    \end{center}

    \begin{flushleft}
    
    {\authors\par}

    \end{flushleft}

}\makeatother

\newcommand{\authors}{
\textbf{Primary author:} Abigail R. Azari, Space Sciences Lab, University of California, Berkeley \\ 
\vspace{0.2 cm}
\textbf{Co-authors\footnote[2]{We would like to recognize the extraordinary effort which this decadal has taken and the members of our community who were unable to participate in this work. We would also like to acknowledge conversations with additional white paper teams on data management, automation, and other machine learning relevant contributions and we encourage you to review these data science relevant papers submitted to the survey. Part of this research was carried out at the Jet Propulsion Laboratory, California Institute of Technology, under a contract with the National Aeronautics and Space Administration.}:} John B. Biersteker$^{1}$, 
Ryan M. Dewey$^{2}$, 
Gary Doran$^{3}$, 
Emily J. Forsberg$^{4}$, 
Camilla D. K. Harris$^{2}$, 
Hannah R. Kerner$^{5}$, 
Katherine A. Skinner$^{6}$, 
Andy W. Smith$^{7}$ \\
\emph{Secondary Contributors}:
Rashied Amini$^{3}$,
Saverio Cambioni$^{8}$, 
Victoria Da Poian$^{9}$,
Tadhg M. Garton$^{10}$,
Michael D. Himes$^{11}$,
Sarah Millholland$^{12}$,
Suranga Ruhunusiri$^{13}$

\vspace{0.5 cm}

\textbf{Co-signers:} 
Andrew M. Annex$^{14}$, 
K.-Michael Aye$^{15}$,
Matthew R. Argall$^{16}$,
Marissa E. Cameron$^{3}$, 
Xin Cao$^{13}$,
Frank Crary$^{15}$,
Daniel J. Crichton$^{3}$,
Matthew Cross$^{17}$,
Mario D'Amore$^{18}$,
Julien de Wit$^{1}$,
Raymond Francis$^{3}$,
Jörn Helbert$^{18}$,
Don R. Hood$^{19}$,
Catriona M. Jackman$^{20}$,
Anthony Lagain$^{21}$,
Raven E. Larson$^{15}$,
Chuhong Mai$^{22}$, 
Lukas Mandrake$^{3}$,
Andrew Needham$^{23}$,
Victor Pankratius$^{1}$, 
James Parr$^{24}$,
Louise M. Prockter$^{25}$, 
Fernando Pérez$^{26}$,
Jani Radebaugh$^{27}$,
Bethany P. Theiling$^{9}$,
Matthew Tiscareno$^{28}$,  
Andrew Vanderburg$^{29}$,
Jon Vandegriff$^{30}$,
Josh Vander Hook$^{3}$,
Indhu Varatharajan$^{18}$,
Kiri Wagstaff$^{3}$,
Ingo Waldmann$^{31}$

\begin{multicols}{2}
\textbf{Affiliations:} \\
$^{1}$~Massachusetts Institute of Technology \\
$^{2}$~University of Michigan \\
$^{3}$~Jet Propulsion Laboratory, California Institute of Technology \\
$^{4}$~University of Idaho \\
$^{5}$~University of Maryland at College Park \\
$^{6}$~Georgia Institute of Technology \\ 
$^{7}$~Mullard Space Science Laboratory, University College London (UCL)\\
$^{8}$~Lunar and Planetary Laboratory, University of Arizona \\
$^{9}$~NASA Goddard Space Flight Center \\
$^{10}$~University of Southampton \\
$^{11}$~University of Central Florida \\
$^{12}$~Princeton University \\
$^{13}$~University of Iowa \\
$^{14}$~Johns Hopkins University (JHU)\\
$^{15}$~Laboratory for Atmospheric and Space Physics, University of Colorado \\
\vspace{0.5ex}
$^{16}$~Space~Science~Center, University~of~New~Hampshire \\
$^{17}$~Institute for Earth and Space Exploration, Western University \\
$^{18}$~German Aerospace Center (DLR)\\
$^{19}$~Texas A\&M University \\
$^{20}$~Dublin Institute for Advanced Studies \\
$^{21}$~Space Science and Technology Centre, Curtin University\\
$^{22}$~Arizona State University \\
$^{23}$~Jacobs, NASA Johnson Space Center \\
$^{24}$~Frontier Development Laboratory \\
$^{25}$~Universities Space Research Association, Lunar and Planetary Institute \\
$^{26}$~University of California, Berkeley, Lawrence Berkeley National Laboratory \\
$^{27}$~Brigham Young University \\
$^{28}$~SETI Institute \\
$^{29}$~University of Wisconsin, Madison \\
$^{30}$~JHU Applied Physics Laboratory\\
$^{31}$~Centre for Space Exochemistry Data, UCL\\
\end{multicols}
}

\begin{document}

\thispagestyle{empty}
\maketitle
\newpage
\setcounter{page}{1}

\section{Introduction}
\label{Intro}
\vspace{-2ex}
In the last decade, machine learning has emerged as one of the most promising complements to traditional data analysis and modeling methods in scientific fields. In one of the most profound examples, the first image of a black hole was captured by applying a machine learning algorithm to petabytes of data collected from eight telescopes \cite{Bouman2016}. 
Since planetary science's last decadal survey, the use of machine learning has increased in each division of NASA’s Science Mission Directorate (SMD). However, even though the number of planetary science publications involving machine learning has grown exponentially over the last ten years, it lags in both percent share and growth rate compared to heliophysics, astrophysics, and Earth science (Figure 1). In this white paper, we assert that planetary science, similar to related disciplines, has the opportunity to leverage machine learning methods for scientific advancement in our field.

\begin{wrapfigure}{R}{0.6\textwidth}
    \begin{center}
    \vspace{-30pt}
    \includegraphics[width=0.6\textwidth]{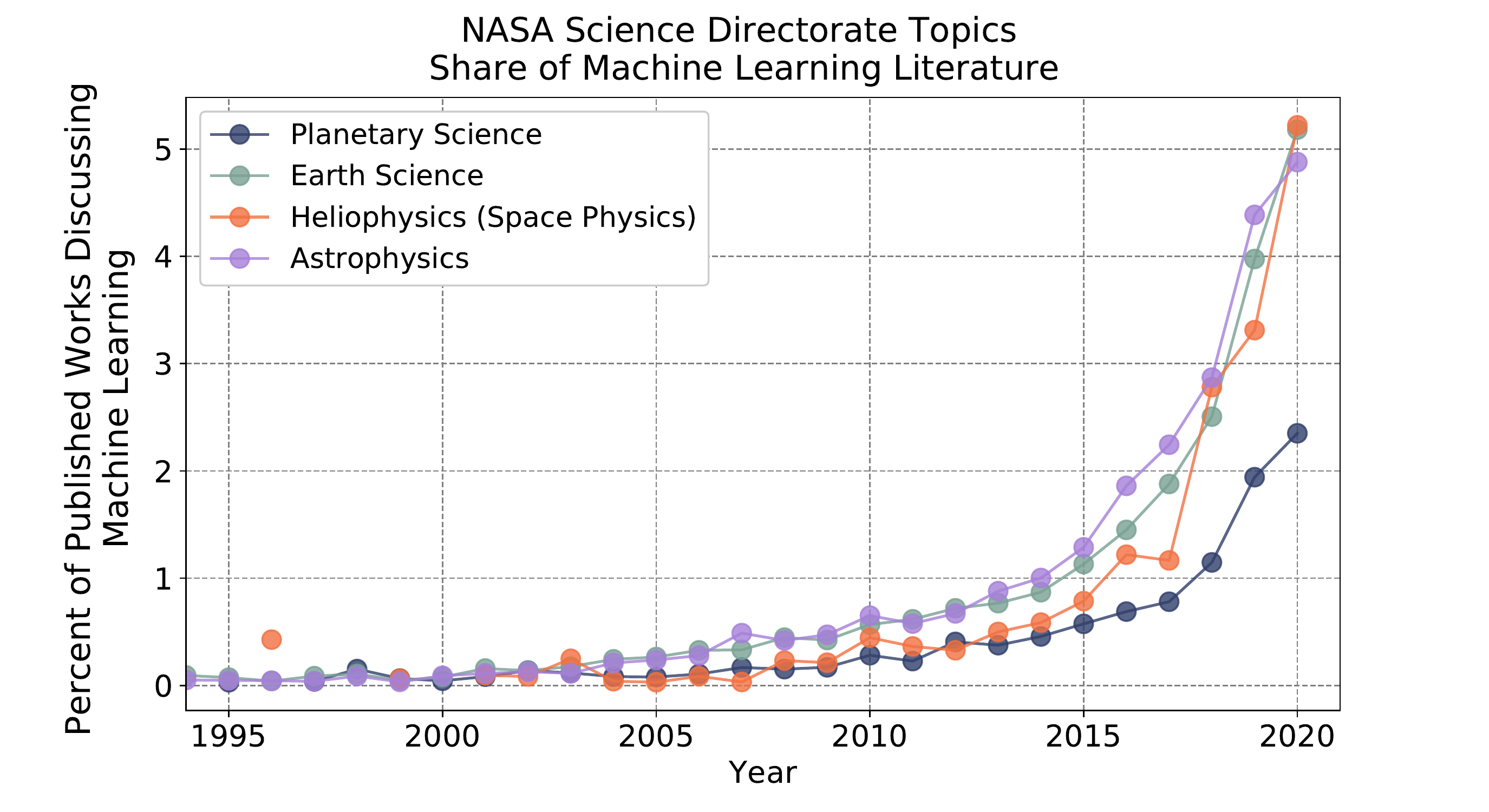}
    \vspace{-35pt}
    \end{center}
    \caption{Yearly trends of machine learning across NASA SMD topics as a percentage of published literature. For comparison, in 2019 the term `modeling' had near identical occurrence rates within publications in both planetary science and Earth science ($>$50$\%$). Data from \href{https://www.scopus.com/}{Elsevier's Scopus} publication database.}
    \label{fig:Status}
    \vspace{-10pt}
\end{wrapfigure}

Planetary science is rapidly transforming into a data rich field with missions in the next ten years set to collect larger and more detailed datasets than ever before. For example, in the fourteen years orbiting Mars, the Mars Reconnaissance Orbiter has already returned more data than all other interplanetary missions combined, including missions currently active (see \href{https://mars.nasa.gov/resources/7741/mars-reconnaissance-orbiter-by-the-numbers/}{MRO By The Numbers}). These missions increase the total data available to our community and increase the rate at which we receive new observations. This can pose challenges to data analysis efforts that machine learning can help address. For example, the Transiting Exoplanet Survey Satellite (TESS) generates ${\sim}1$ million new light curves every month, pushing the exoplanet community toward automated candidate identification \cite{Yu2019}.  Integrating machine learning into planetary science has the potential to enhance current endeavors and lead to new discoveries in our field.

The broad applicability of data science methods is considered a new branch of scientific investigation \cite{Hey2009}. Data from planetary science missions and the field's interest in interpretable models and reproducible science pose challenges to immediate application of off-the-shelf machine learning methods \cite[e.g.,][]{Azari2020}. However, these challenges are shared by other fields, including Earth science~\cite{Karpatne_2019}. We can learn from demonstrated funding programs and researchers in sibling disciplines to adopt field-specific machine learning methods. These efforts should embrace interdisciplinary partnerships across multiple application fields and industry to address unique challenges in planetary science as they manifest in the context of machine learning. Our vision for the next decade is one rich with applications of machine learning at every stage of planetary missions. We envision machine learning enhancing current activities by operating alongside established methods. 

\begin{center}
\textbf{In recognition of the novelty of these methods to the field, a concerted effort is needed to ensure integration and applicability of machine learning methods in planetary science.}
\end{center}

\begin{wrapfigure}{R}{0.6\textwidth}
\vspace{-2pt}
\begin{mybox}[title=What is Machine Learning?]

Machine learning (ML) is a type of artificial intelligence that refers to computer algorithms that improve (or learn) through inclusion of data. ML methods are a highly flexible class of algorithms designed to adjust their parameters to the structure of the data they have access to and produce models that make predictions. This flexibility makes them suitable for various uses and the scientific community is still exploring and discovering novel ways in which they can become part of our toolbox. In the context of this white paper, ML ranges in complexity from linear regression to neural networks, with applications ranging from image classification to pattern recognition. Within this white paper we use data science as an umbrella term for methods that take advantage of large data sets, such as machine learning.

\end{mybox}
\vspace{-10pt}
\end{wrapfigure}

The purpose of this white paper is two-fold. First, to introduce and demonstrate to the reader the applicability of these methods at every stage of a mission (\autoref{Lifecycle}). Second, to offer recommendations to ensure the next decade of planetary science capitalizes on the benefits of machine learning for scientific advancement (\autoref{Reccs}). \textbf{To support these applications we recommend a sustained commitment to the ecosystem of machine learning in planetary science, independent of individual mission funding.}

\vspace{-2ex}
\section{Machine Learning at Every Stage of the Mission Lifecycle}
\label{Lifecycle}
\vspace{-2ex}
\subsection{Mission Development} \label{sec:mission-development}
\vspace{-1ex}

\noindent Mission development requires surveys of existing datasets and the development of models to translate science goals into discrete engineering requirements. Data science techniques are already informing the design of missions currently under development. In this section we highlight several examples of how ML can be incorporated to efficiently develop future missions.

\vspace{0.2em}    
\noindent \textbf{Planning trajectories}\label{dev-trajectory} for missions is critical for meeting requirements. Trajectory design is constrained by vehicle configuration and fuel capacity, resulting in a non-trivial problem that can be computationally expensive~\cite{Mereta_2017}. ML can be used to improve accuracy and reduce computational cost of optimal trajectory design. Recent work~\cite{Izzo_2019} has shown the potential for ML to address current challenges by demonstrating the feasibility of training a deep neural network on the ground to enable real-time onboard spacecraft guidance.

\vspace{0.2em}    
\noindent \textbf{Earth and laboratory analog studies} can be combined with machine learning to develop mission-like training datasets for missions exploring new environments. This style of transfer learning, in which a trained machine learning model is used on a different dataset, could offer significant advantages from the bevy of Earth data we have toward mission planning. 

\vspace{0.2em}    
\noindent \textbf{Initial landing site selection}\label{dev-landing} is labor intensive and can be subject to human errors, with gigabytes of terrain maps being labeled by hand and locations evaluated with respect to predicted landing ellipses. ML and data-driven techniques can efficiently provide high-level classification of surface terrain across large areas to automate preliminary selection of potential sites. Recent work to aid in landing site selection of the Mars 2020 mission~\cite{Ono_2016} leveraged ML to enable automated classification of orbital images by terrain type, allowing scientists to focus on choosing landing sites with the highest potential for scientific return.

\vspace{0.2em}    

\noindent \textbf{Evaluation of mission architecture}\label{dev-predict} and instrument design against scientific requirements cycles through design, testing, and improvement. Data science techniques can improve and accelerate this cycle through robust uncertainty propagation and surrogate models that approximate the results of high-fidelity simulations. For example, during hardware testing, an iterative approach using Bayesian inference can improve model validation and reduce uncertainty in instrument performance \cite{Stout2015}. Probabilistic models of individual components can be combined into estimates for the performance envelope of an entire system, resulting in more reliable benchmarking against design requirements early in the design cycle \cite{Uebelhart2005}. Probabilistic models combined with models of the physical system can generate simulated datasets that data pipelines and analysis techniques can be tested against for satisfaction of requirements.

These examples show how data science methods impact the development of missions. In the next decade we expect machine learning techniques to improve mission designs by enabling the use of larger datasets and enhancing mission development.

\vspace{-2ex}

\vspace{-0.5ex}
\subsection{Mission Operations} \label{Operations}
\vspace{-1ex}

\noindent Machine learning shows promise for increasing scientific return from planetary missions by addressing limitations in science operations and data return. While planetary missions have increasingly generated and returned more data to Earth over the last several decades, physical limitations on communication bandwidth and latency combined with an increasing number of missions that must share Deep Space Network resources mean that much less data can be returned compared to Earth-orbiting missions. Data science techniques can automate or assist with science operations. The use of ML in tandem with onboard planning enables new mission architectures and exploration of high science value, high risk destinations. In this section, we describe opportunities for integrating ML into operations onboard spacecraft and on the ground.

\vspace{0.2em}
\noindent \textbf{Automated content-based alerts} can accelerate science planning by notifying scientists of interesting observations. Scientists might have as little as 5 hours to decide which commands to send to the spacecraft at the next communication opportunity \cite{Wilson2017}. Using automated classification, science planners could filter new observations to review only those containing features of interest, or could ``subscribe'' to be notified when an onboard algorithm detects such features. Current examples include implementing a convolutional neural network to classify geologic features and spacecraft hardware within Mars rover and orbital imagery \cite{Wagstaff2018} and automatically generating image captions for content-based search or notification capabilities \cite{Ono2019}.  

\begin{wrapfigure}{R}{0.55\textwidth}
\begin{center}
\vspace{-25pt}
\includegraphics[width=0.55\textwidth]{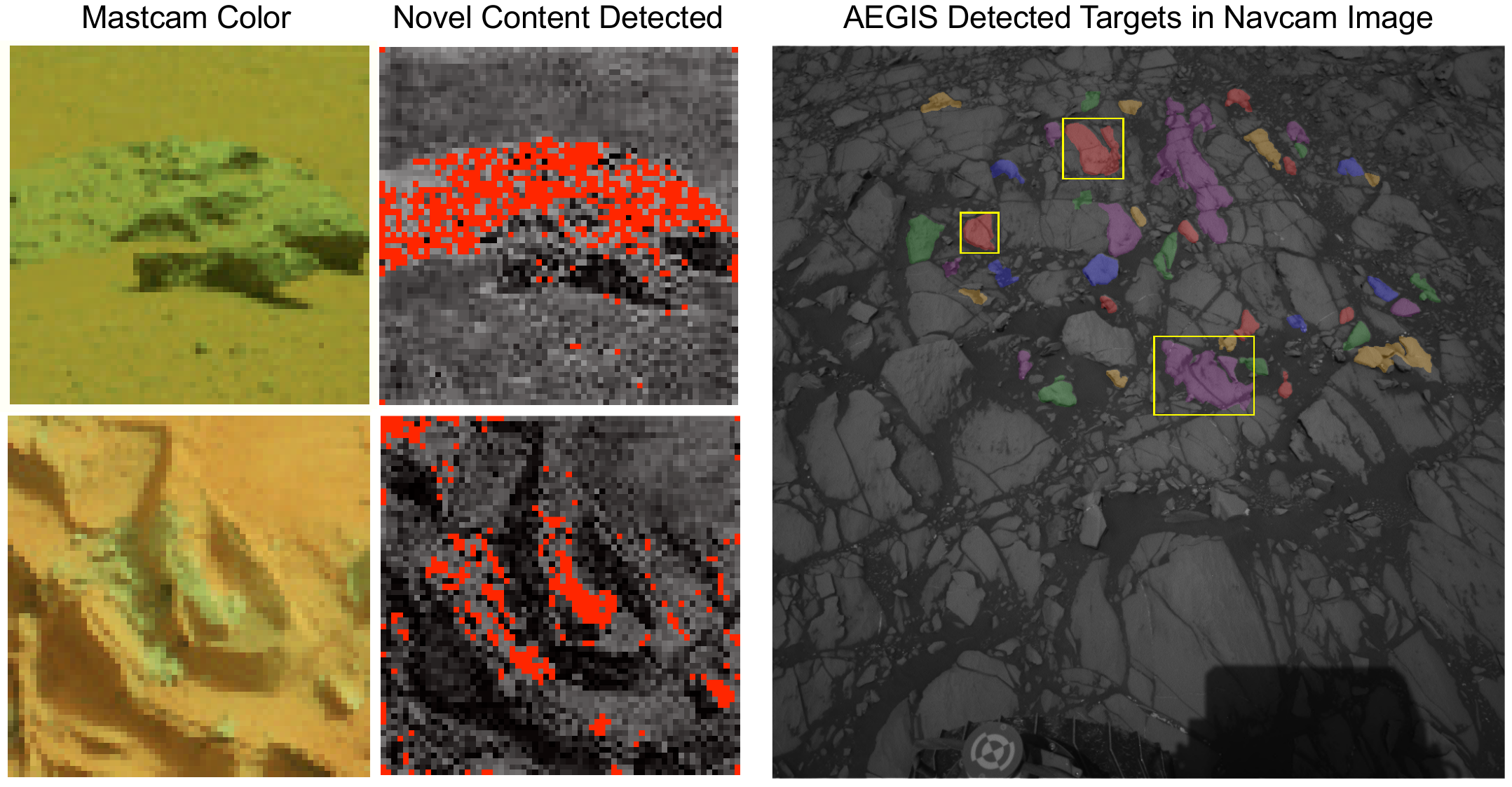}
\vspace{-30pt}
\end{center}
\caption{Examples of automated feature detection on Mars with Mars Science Laboratory. Left: Mastcam images of a meteorite (top) and light-toned material (bottom) corresponding to red pixels detected by a ML algorithm \cite{Kerner2019}. Right: Navcam image of the Martian surface highlighting targets for follow-up analysis from the Autonomous Exploration for Gathering Increased Science (AEGIS) \cite{Francis17aegis}.}
\label{fig:MarsCam}
\vspace{-18pt}
\end{wrapfigure}

\vspace{0.2em}    
\noindent \textbf{Prioritization of novel observations} can also help focus scientists' attention on unusual phenomena and respond with any follow-up observations while there is still an opportunity to do so. Applications have prioritized the order of spectra shown to an expert, improving the accuracy of spectrum classification \cite{Wagstaff2013} and have identified novel geology in multi-spectral images to flag observations for mission team members \cite{Kerner2020b}. See \autoref{fig:MarsCam} (left) for a successful example on detecting Martian geologic features.

\vspace{0.2em}       
\noindent \textbf{Onboard detection and summarization} (i.e., high-level observation summaries) of features can greatly increase the number of observations containing these phenomena returned from a mission. It is nearly impossible to sequence observations of transient, unpredictable events (e.g., dust devils, storms, plumes) in advance. Searching for phenomena onboard a spacecraft enables downlinking only those candidate observations likely to contain the desired target or returning more data for targets \cite{Wagstaff2019clipper}. This approach has been used on several missions, including the delivery of observation content summaries from EO-1 \cite{Thompson2012autonomous} and the automatic detection of dust devils on Mars Exploration Rover Opportunity \cite{Chien2008watch}. Even imperfect algorithms will increase the fraction of valuable observations in the downlink.

\vspace{0.2em}
\noindent \textbf{Autonomous targeting} can recover mission time spent waiting for ground-in-the-loop targeting decisions. Such methods are becoming essential for certain operations. For example, the AEGIS system (\autoref{fig:MarsCam}), used to identify geologic targets for follow-up observations onboard Mars Exploration Rovers and Mars Science Laboratory, has increased observations by $\sim$28\% per day ~\cite{Francis17aegis}. 

Onboard autonomy will be increasingly important for missions at far distances (icy moons and icy giants), with multiple spacecraft, and with short duration and limited communication opportunities (Europa and Venus landers). Applications of autonomy are extremely valuable, but face challenges in hardware and software design and in producing explainable and interpretable results. Long-term success will require addressing these challenges (\autoref{Reccs}), but promises to improve the scientific return of planetary missions.

\vspace{-2ex}
\vspace{-0.5ex}
\subsection{Analyses} \label{sec:analyses}
\vspace{-1ex}
\noindent Statistical techniques have long been used to analyze planetary data to advance scientific understanding. Recent increases in data volume returned by planetary missions have expanded the applicability of data science techniques. We present a high level overview of leading edge applications that these methods and data volumes enable. 

\vspace{0.2em}
\noindent \textbf{Detection, characterization, and classification} with ML can systematically and autonomously identify features of interest and distinguish real objects and events from false positives. Applications range from standardizing and identifying rare plasma intensifications in a time series \cite{Azari2018} to vetting large datasets for exoplanet candidate identification \cite{Millholland2017}. Compared to compiling such events or features by eye, ML can be employed as a tool to find and reduce bias while improving reproducibility for scientific analyses.

\begin{wrapfigure}{R}{0.4\textwidth}
    \begin{center}
    \vspace{-40pt}
    \includegraphics[width=0.4\textwidth]{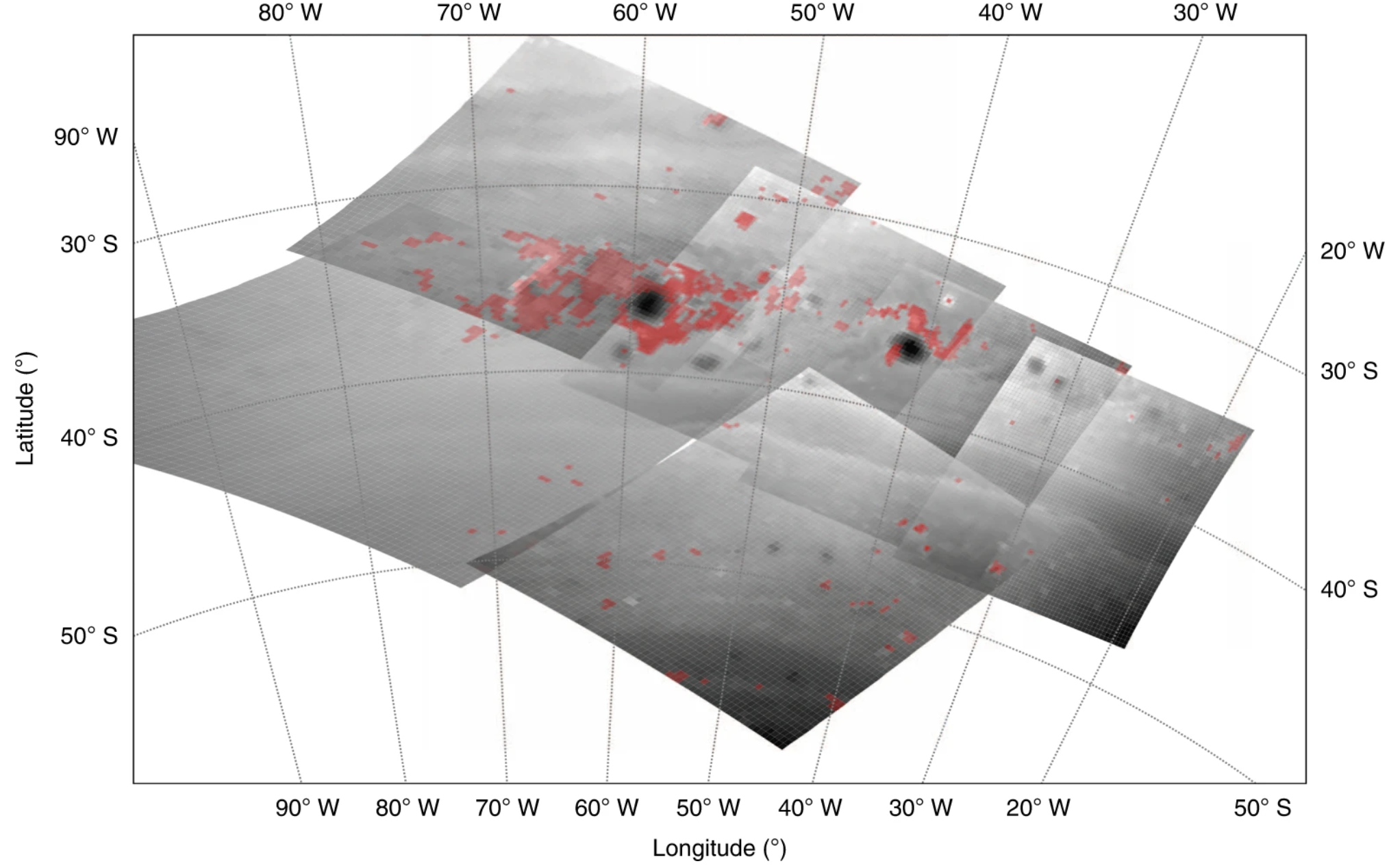}
    \vspace{-35pt}
    \end{center}
    \caption{Stormy regions (red) in Saturn's clouds detected by PlanetNet. Reprinted from ~\cite{Waldmann2019} by permission from 
    \href{https://www.nature.com/natastron/}{Nature Astronomy}.
    }
    \label{fig:Waldmann2019}
    \vspace{-35pt}
\end{wrapfigure}

\vspace{0.2em}
\noindent\textbf{Grouping features (clustering)} into categories with statistics-based assumptions has the potential to be a major method leading to groundbreaking findings in planetary science. Such techniques have been implemented in Bayesian clustering for mineral identification on Mars \cite{Dundar2016}. ML also offers the opportunity to develop techniques that can be quickly and feasibly applied to much larger data volumes. For example, the development of PlanetNet, a tool to ``quickly and accurately map spatial and spectral features across large, heterogeneous areas'' illustrated in \autoref{fig:Waldmann2019}, uses a dataset that is too large for manual identification \cite{Waldmann2019}. 

\vspace{0.2em}    
\noindent\textbf{Resolving (and integrating across) multi-dimensional non-linear trends} is of great interest in studying planetary environments. Traditional statistical analysis struggles to account for these multivariate trends, particularly when the complete parameter space has not been thoroughly sampled. Data science allows for the identification and description of simultaneous, multivariate, and nonlinear trends; for example using temporal variability to estimate spatial scales \cite{Smith2018}.

\vspace{0.2em}
\noindent \textbf{Describing large physical systems} with spatiotemporal varying datasets can be challenging. Planetary missions can only capture small snapshots of large-scale systems. Statistical techniques can use these snapshots to construct global descriptions of the system. This can be done, for example, by combining observations into useful proxies \cite{Ruhunusiri_2018}. 

\vspace{0.2em}
\noindent \textbf{ML integration in physical models} can build on the physics-based modeling already embraced by the field. Large-scale physical models have been a mainstay for interpreting planetary science datasets for decades and ML techniques are quickly gaining use in enhancing physics-based model performance and returns \cite{Cambioni_2019}.

Advanced statistical techniques are adding a new branch of methods with which to engage in data-driven scientific pursuit. Similar to other mission stage applications, interpretability of these methods’ results is an active challenge. To fully enable ML applications, it is therefore of critical importance to develop these techniques within planetary science.

\vspace{-2ex}
\section{Recommendations}
\label{Reccs}
\vspace{-2ex}
To reach our future vision of the field, we have crafted ten recommendations for integration of machine learning into planetary science for the next decade.

\vspace{-1ex}
\begin{enumerate}[wide, labelwidth=!, labelindent=0pt, label=\textbf{\arabic*}.]
\setlength\itemsep{-0.2em}

\item{\textbf{Invest in post-mission and non-mission related funding and activities.}} Due to the structure of the NASA Planetary Science Division, funding opportunities are often tied to missions or demonstrable scientific goals. Because ML techniques are best suited to larger datasets, we expect many ML proposals will focus on missions considered retired or ended. Increased funding for post-mission studies would assist in these applications. Beyond missions, research is needed for how ML techniques will address specific challenges in planetary science data. ML applications to physical problems are still a major research area. Recent findings highlight the need for incorporating domain knowledge to increase interpretability of ML \cite{Azari2020} and propose a machine assisted pipeline in scientific discovery \cite{Pankratius2016}. Basic investigations into domain-specific concerns of ML techniques are needed and are not funded by missions.

\item{\textbf{Expand programs for data analysis.}}
ML applications currently fit into data analysis programs, such as the existing NASA Planetary Data Archiving, Restoration, and Tools and Data Analysis Programs. We recommend a review of these programs for inclusion of ML.

\item\textbf{{Increase support for planetary data infrastructures.}} The NASA Planetary Data System (PDS) forms the backbone of NASA's planetary data.
We recommend increased discussions for PDS and other data and tool hubs (e.g. \href{https://www.lpi.usra.edu/mapsit/}{MAPSIT}) to support ML applications. 

\item{\textbf{Invest in onboard software and hardware.}} 
A maturation pipeline is needed to mitigate risk of including instruments and spacecraft capable of onboard data processing in mission proposals. We recommend including ML in \href{https://astrobiology.nasa.gov/research/astrobiology-at-nasa/picasso-and-matisse/}{PICASSO and MatISSE} or developing new programs for increasing technology readiness levels of onboard software and automation. 

\item{\textbf{Invest in the continued development of data return facilities.}} For ML users interested in data from missions traveling far from Earth (like the upcoming Dragonfly and Europa Clipper missions), data return and the cost of transmission limit data sizes. In addition to onboard solutions (see previous item), we support investments for updates to the Deep Space Network and other solutions for higher data return from the outer planets.

\item{\textbf{Review changing demands for computational resources on the ground.}} Greater usage of NASA's high performance computing facilities is expected as ML applications increase the demand on computational resources. We recommend that these facilities receive continued support.

\item{\textbf{Invest in education.}} Educational resources for new users of ML methods are essential. Research programs like the \href{https://frontierdevelopmentlab.org/}{Frontier Development Laboratory} work to integrate artificial intelligence with NASA research and provide training for students and early career professionals. We recommend expansion of these programs and investigation into models that would educate larger groups (see \href{https://www2.cisl.ucar.edu/events/summer-school/ai4ess/2020/artificial-intelligence-earth-system-science-ai4ess-summer-school}{AI4ESS} for a potential model). On longer timelines we see planetary science degrees incorporating data rich courses \cite[see][for necessity of such courses in education]{NASEM_2018} and hiring the staff required to teach such courses (see \href{https://www.nsf.gov/funding/pgm_summ.jsp?pims_id=504953}{NSF FDSS} for an program that encouraged investment through faculty sponsorship).  

\item{\textbf{Support community groups and early career researchers.}} While the community of ML users in planetary science is growing, in our observation it is less cohesive than other subsets of the field and is primarily composed of early career participants. On this white paper for example, all of the authors are early career professionals or students. Supporting early career researchers and groups like \href{https://www.openplanetary.org/}{OpenPlanetary} that focus on planetary data access will help connect researchers. The NSF EarthCube \href{https://is-geo.org/}{Intelligent Systems for Geosciences} provides an ideal example.

\item{\textbf{Actively support underrepresented members of our community.}} 
Geoscience and computer science are source fields for ML in planetary science. These fields, respectively, are the least racially diverse of all STEM fields \cite{Dutt2020} and have some of the lowest shares of women degree recipients with declining participation in recent years \cite{NSF2019}. Opportunities must actively include the participation, experiences, and leadership of underrepresented members of our communities or they will reinforce existing inequalities and propagate injustices into the next decade. 

\item{\textbf{Support interdisciplinary work.}} For the continued success and growth of ML in domain applications, scientific fields must engage in interdisciplinary work \cite{Ebert2019}, such as with ML industry researchers and physical scientists of other fields. This furthers Strategy 1.3 in the NASA SMD 2020-2024 \href{https://science.nasa.gov/science-red/s3fs-public/atoms/files/2020-2024_Science-TAGGED.pdf}{strategy plan} to ``Advance discovery in emerging fields by identifying and exploiting interdisciplinary opportunities'' \cite{NASA2020}. 
\end{enumerate}

\vspace{-2ex} 
\begin{mybox}[title=A Single Solution: `SpaceCube']

Our recommendations represent different aspects of a single target: the sustained commitment to people, applications, and infrastructure outside of mission funding. We have looked across fields for examples for our future, referencing the joint program between the NSF Directorate for Geosciences and the Division of Advanced Cyberinfrastructure called \href{https://www.earthcube.org/}{NSF EarthCube}. For Earth research, EarthCube envisions almost every aspect of our recommendations. EarthCube funded projects have supported community groups, worked to create data infrastructures, and defined new research areas. 
Given NASA's excellence in planetary research, and NSF's EarthCube program, an NSF-NASA partnership would further our vision for the field to engage in purposeful work in  machine learning for planetary science.
\vspace{-2ex}
\begin{center}
\textbf{We propose a `SpaceCube' partnership between NASA and NSF to target areas of shared advancement to address these recommendations.}
\end{center}
\end{mybox}
\vspace{-2ex}

\vspace{-2ex}
\section{Conclusions}
\label{Conc}
\vspace{-2ex}
The future of planetary science is exciting, with each new discovery changing our understanding of the solar system and beyond. Machine learning and data science are poised to advance these discoveries and add value at each step of the mission lifecycle. In this white paper we have envisioned a data rich future for our field achieved by integrating these methods. For the full incorporation of machine learning into planetary science, investigations into domain-specific applications for our field must be supported. The successes of adjacent fields have informed our recommendations, which will enable planetary science to join them in benefiting from machine learning techniques for scientific insight. The NSF EarthCube program exemplifies many of our recommendations; but for Earth science. As such, we urge the survey to consider cross-agency collaborations to leverage current activities outside of NASA.  The \href{https://solarsystem.nasa.gov/resources/598/vision-and-voyages-for-planetary-science-in-the-decade-2013-2022/}{last decadal survey} stated that planetary missions have benefited from ``laboratory studies, theoretical studies, and modeling'' \cite{LastDecadal}. 
We look forward to seeing data science and machine learning methods included in the next decade.

\printbibliography

\end{document}